\documentclass[twocolumn]{aastex631}
\usepackage{amsmath}
\usepackage{CJK}

\begin{document}
\begin{CJK*}{UTF8}{gbsn}

\title{Detailed radial scale height profile of dust grains as probed by dust self-scattering in HL Tau}

\correspondingauthor{Haifeng Yang}
\email{hfyang@zju.edu.cn}

\author[0000-0002-8537-6669]{Haifeng Yang (杨海峰)}
\altaffiliation{ZJU Tang Scholar}
\affiliation{Institute for Astronomy, School of Physics, Zhejiang University, 886 Yuhangtang Road,  Hangzhou 310027, China}
\affiliation{Center for Cosmology and Computational Astrophysics, Institute for Advanced Study in Physics, Zhejiang University, Hangzhou 310027, China}

\author{Ian W. Stephens}
\affiliation{Department of Earth, Environment, and Physics, Worcester State University, Worcester, MA 01602, USA}

\author[0000-0001-7233-4171]{Zhe-Yu Daniel Lin}
\affiliation{Department of Astronomy, University of Virginia, Charlottesville, VA 22903, USA}

\author[0000-0001-5811-0454]{Manuel Fern\'andez-L\'opez}
\affiliation{Instituto Argentino de Radioastronom\'ia (CCT-La Plata, CONICET; UNLP; CICPBA), C.C. No. 5, 1894, Villa Elisa, Buenos Aires, Argentina}
\affiliation{Facultad de Ciencias Astron\'omicas y Geof\'isicas, Universidad Nacional de La Plata, Paseo del Bosque S/N, B1900FWA La Plata, Argentina}

\author[0000-0002-7402-6487]{Zhi-Yun Li}
\affiliation{Department of Astronomy, University of Virginia, Charlottesville, VA 22903, USA}

\author[0000-0002-4540-6587]{Leslie W. Looney}
\affiliation{Department of Astronomy, University of Illinois, 1002 West Green Street, Urbana, IL 61801, USA}

\author{Rachel Harrison}
\affiliation{School of Physics and Astronomy, Monash University, Clayton VIC 3800, Australia}



\begin{abstract}
The vertical settling of dust grains in a circumstellar disk, characterized by their scale height, is a pivotal process in the formation of planets. This study offers in-depth analysis and modeling of the radial scale height profile of dust grains in the HL Tau system, leveraging high-resolution polarization observations. We resolve the inner disk's polarization, revealing a significant near-far side asymmetry, with the near side being markedly brighter than the far side in polarized intensity. This asymmetry is attributed to a geometrically thick inner dust disk, suggesting a large aspect ratio of $H/R \ge 0.15$. The first ring at 20 au exhibits an azimuthal contrast, with polarization enhanced along the minor axis, indicating a moderately thick dust ring with $H/R \approx 0.1$. The absence of the near-far side asymmetry at larger scales implies a thin dust layer, with $H/R < 0.05$. Taken together, these findings depict a disk with a turbulent inner region and a settled outer disk, requiring a variable turbulence model with $\alpha$ increasing from $10^{-5}$ at 100 au to $10^{-2.5}$ at 20 au. This research sheds light on dust settling and turbulence levels within protoplanetary disks, providing valuable insights into the mechanisms of planet formation.

\end{abstract}



\section{Introduction} \label{sec:intro}




Protoplanetary disks are the cradles of planetary systems, offering a unique window into the early stages of planet formation \citep{Armitage2011,Andrews2020}. 
Under the core accretion paradigm, interstellar dust grains within these disks grow through coagulation, eventually forming planetesimals and, ultimately, planets \citep{Pollack1996}. A critical step in this process is to increase the concentration of dust grains, which is facilitated by the vertical settling of dust grains towards the midplane of the disk \citep{Chiang2010}. 
In addition, the vertical structure of protoplanetary disks also holds key insights into the physical
environment, such as the turbulence level in the disk where planet formation takes place.
Although numerous studies have investigated the vertical scale height of dust grains in protoplanetary disks observationally, most of these have concentrated on edge-on systems, providing a clear view of the disk's vertical structure \citep{Lee2017,Sakai2017,Ohashi2022,Michel2022,Villenave2020,Villenave2022,Lin2023}. 

The scattering-induced polarization offers a novel method for investigating the vertical dust scale heights in moderately inclined protoplanetary disks. This approach represents an exciting frontier in astrophysics, significantly advanced by the high-resolution polarization capabilities of the Atacama Large Millimeter/submillimeter Array (ALMA). Observations from ALMA have suggested the presence of 100 $\mu$m dust grains within these disks, providing valuable insight into the grain growth process (see, e.g., \citealt{Kataoka2015,Yang2016a,Hull2018}).
Importantly, polarization arising from dust scattering is highly sensitive to the dust's vertical distribution.
Optically and geometrically thick disks, for example, exhibit a characteristic near-far side asymmetry in polarization, directly linked to their vertical structure \citep{Yang2017}. Furthermore, scattering-induced polarization is influenced by the anisotropy of the radiation reaching the dust grains, which in turn depends strongly on the optical depth through the midplane, a factor indirectly governed by the dust's vertical scale height. The interaction between inclination-induced polarization and this radiation anisotropy thus provides robust constraints on the dust scale height in disks \citep{Ohashi2019,Lin2023,Yang2024}.

In this study, we focus on the HL Tau system, a well-studied protoplanetary disk that has been extensively observed with ALMA. 
One of the ALMA early science campaigns yielded detailed ring and gap structures in this system, which has revolutionized the field of planet formation \citep{ALMA2015}. 
It is also the first Class I or T Tauri disk that was resolved with polarization observations, showing roughly uniform
polarization, opening a new field of study in scattering-induced polarization \citep{stephens2014,Yang2016a,Kataoka2016HLTau}.
The new high-resolution polarization observations presented by \cite{Stephens2023} 
reveal that the polarization pattern in the gap is also in azimuthal configuration, similar to what is observed in ALMA Band 3 (3 mm) \citep{Kataoka2015, Stephens2017}. 
So there exist a transition of polarization patterns from uniform patterns to azimuthal patterns, both in wavelength domains from 0.87 to 3 mm, and in spatial domain from rings to gaps. 
Qualitatively, the transition can be explained by a change in optical depth \citep{Stephens2023,Lin2024a}.
In addition, the high-resolution polarization observations also
provide a unique opportunity to probe the dust grain properties and their radial scale height profile,
especially at small scales that are not resolved in earlier polarization observations.
By analyzing the dust self-scattering signal, we can infer the detailed structure of the disk and the processes that govern dust settling, which is the goal of this paper.

This paper is structured as follows. In Section~\ref{sec:observation}, we present the transformed data, the target of our modeling efforts in this work. In Section~\ref{sec:model}, we model the high-resolution polarization from dust self-scattering, focusing on three different scales in three different subsections.
In Section~\ref{sec:discussion}, we discuss the insights of our results on the turbulence levels, and present
the final model with residuals. 
We conclude our work in Section~\ref{sec:summary}.

\section{Observation}
\label{sec:observation}
The high-resolution polarization observations used in this paper are from the ALMA project 2019.1.01051.S
(PI: I. Stephens), combined in part with the ALMA project 2016.1.00115.S (PI: I. Stephens). 
These projects observe HL Tau in Full Stokes polarization at ALMA Band 7 (870 $\mu$m).
The details of the observations and the data reductions are presented in \citep{Stephens2023}.
The synthesized beam is $35.5\, \mathrm{mas}\times 30.5\,\mathrm{mas}$, with a position angle of $22^\circ$, using a robust parameter of 0.5.
This corresponds to a resolution of $5.2$ au at a distance of $147.3$ pc \citep{GaiaDR3}. 
The rms noise level of the Stokes I image is $15.9\rm\, \mu Jy/beam$, and the rms noise levels of the Stokes Q and U
images are the same as $8.4\rm\, \mu Jy/beam$. 
Polarization maps are shown in Figures 1 and 3 of \citep{Stephens2023}.

\begin{figure}[!htp]
    \includegraphics[width=0.5\textwidth]{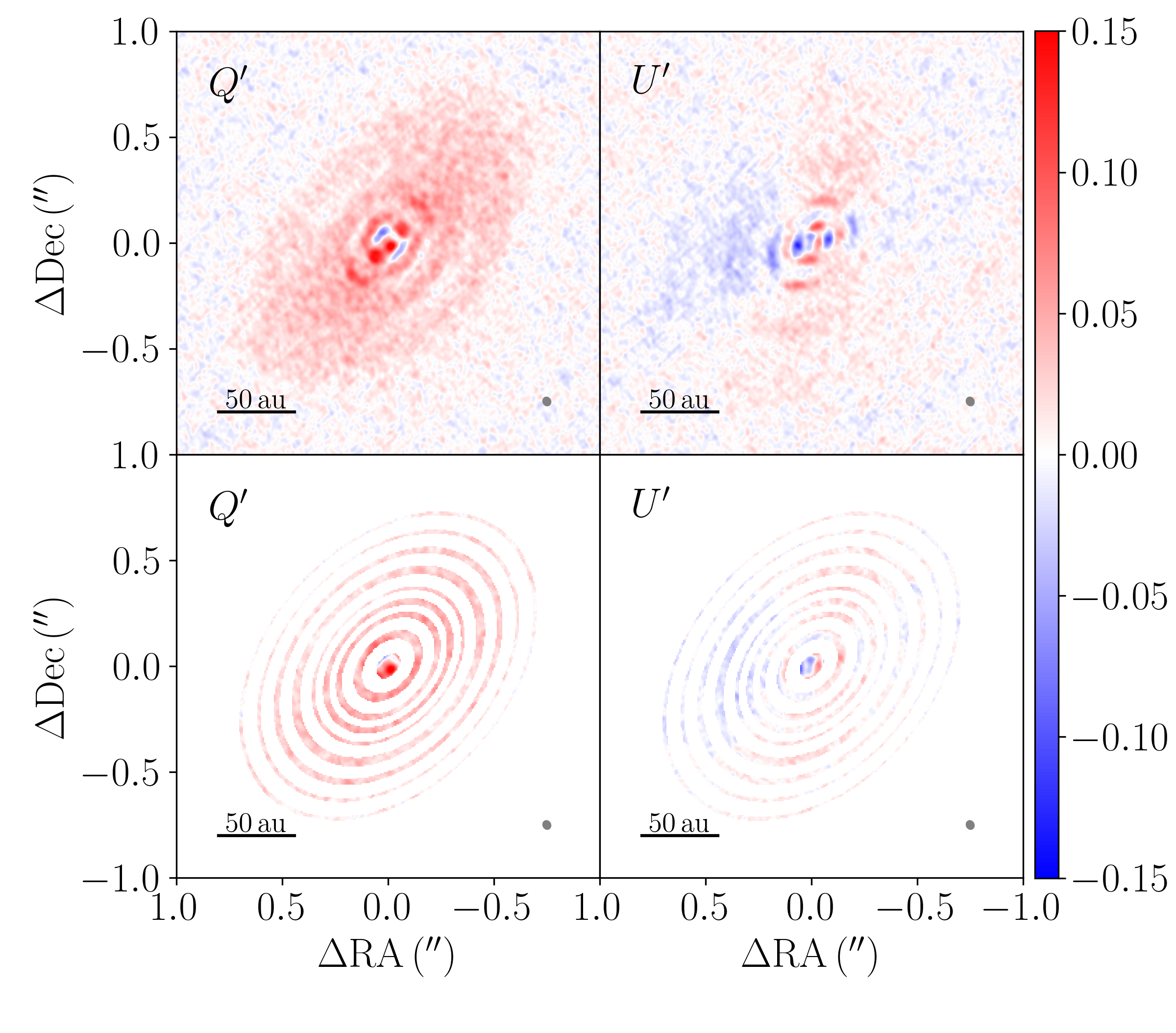}
    \caption{\textbf{Upper panels}: The transformed Stokes parameters $Q'$ and $U'$. 
    The units are $m\rm Jy/beam$ for all panels.
    \textbf{Lower panels}: The masked Stokes $Q'$ and $U'$ that only show data on the rings and the inner disk. The masked $Q'$ is mostly positive and larger than $U'$ in the masked regions.
    The polarization orientation is mostly uniform along the minor axis of the disk, coming from inclination-induced polarization. This is the target of the modeling in this work.}
    \label{fig:image}
\end{figure}

In this work, we aim to study the parts of the high-resolution polarization image that are dominated by self-scattering of dust grains, both in the inner disk and in the rings.
In a moderately inclined disk, like HL Tau, the self-scattering is known to produce uniform polarization patterns along the minor axis \citep{Yang2016a}.
With this in mind, we define Stokes $Q'$ and $U'$ as follows:
\begin{equation}
    \left\{
    \begin{aligned}
        Q'&= Q\cos[2(\mathrm{PA}-90^\circ)]+U\sin[2(\mathrm{PA}-90^\circ)]\\
        U'&= U\cos[2(\mathrm{PA}-90^\circ)]-Q\sin[2(\mathrm{PA}-90^\circ)]
    \end{aligned}
    \right.,
\end{equation}
where $\mathrm{PA}=138.02^\circ$ is the position angle of the disk. 
Under this transformation, the Cartesian coordinates where the Stokes parameters are defined are transformed into a coordinate with the
minor and major axes being the $X$ and $Y$ axes, respectively.
A positive (negative) $Q'$ means polarization along the minor (major) axis.
The transformed Stokes $Q'$ and $U'$ are shown in the upper panels of Figure~\ref{fig:image}.
The location and width of the 8 rings are taken from \cite{Stephens2023} and listed in Table~\ref{tab:daniel_pars}. 

In the lower panels of Figure~\ref{fig:image}, we mask the data so that only rings are shown, as they are the focus of this investigation. 
We can see that in the rings and the inner disk, the polarization is dominated by $Q'$, which means that the polarization is dominated by
uniform polarization patterns along the minor axis of the disk.
In this work, we will focus on modeling the $Q'$ maps.
There are some systematic patterns with peaks slightly above $3\sigma$ in the $U'$ 
map that cannot be explained by self-scattering. They are smaller than the $Q'$ signals at the same radii. 
We will ignore these signals in this work, except for some discussions in Section~\ref{ssec:residuals}.

\section{Model}
\label{sec:model}
To model scattering-induced polarization in the HL Tau system, we set up a disk model, similar to the one adopted by \cite{Stephens2023}. 
For the dust model, we adopt the DSHARP composition \citep{Birnstiel2018}, which
has 0.2 water ice \citep{Warren2008}, 0.3291 astronomical silicate \citep{Draine2003}, 0.0743 troilite \citep{Henning1996}, and 0.3966 refractory organics \citep{Henning1996}, all in mass fractions.
We assume a minimum grain size of $0.1\rm\, \mu m$ and a maximum grain size of $140\rm\, \mu m$.
We fix the dust model, as the aim of this paper is to study the effect of dust vertical distributions on the
detailed structures of scattering-induced polarization. 
We introduce a few modifications to this model in order to better reproduce the Stokes I image with our dust model visually. This paper focuses on comparing polarization at the same radius (for both the near-far side asymmetry and the azimuthal contrast), so the exact choice of radial profiles does not affect our conclusions.

The temperature profile is assumed as a power law profile, $T=T_{10}(R/10\rm\, au)^{-0.65}$, where
$T_{10}=137.5\rm\, K$ is $25\%$ higher than that adopted by \cite{Stephens2023} and the power law index is also steeper.
For the inner disk, we assume:
\begin{equation}
    \tau_{ID}(R)=\tau_{c,0}\left(\frac{R}{R_c}\right)^{-0.5}\exp\left[-\left(\frac{R}{R_c}\right)^{-1.5}\right],
\end{equation}
where $\tau_{c,0}=2.3$ is a characteristic optical depth. In addition, we also add 8 rings, prescribed as
Gaussian profiles in radial direction, as follows:
\begin{equation}
    \tau_i(R) = \tau_{c,i} \exp\left[-\frac{1}{2}\left(\frac{R-C_i}{W_i}\right)^{-2}\right],
\end{equation}
where $\tau_{c,i}$, $C_i$, and $W_i$, are the central optical depth, the central radial location, and the width of the ring, respectively. Their values are tabulated in Table~\ref{tab:daniel_pars}.

\begin{table}[!htp]
    \centering
    \begin{tabular}{|c|cccccccc|}
    \hline
       Ring \#  &  1 &  2 &  3 &  4 &  5 &  6 &   7 &   8 \\
    \hline
       C [au]   & 24 & 39 & 49 & 59 & 73 & 88 & 102 & 116 \\
       $\tau_c$ &  8 &  5 &  5 &  8 &  3 &  8 &   8 &   8 \\
       W [au]   &  4 &  3 &  3 &  2 &  4 &  3 &   2 &   2 \\
    \hline
    \end{tabular}
    \caption{The parameters to define the 8 rings in our model. From top to bottom shows the ring number, central radial location, the central optical depth, and the width of each ring. Taken from \cite{Stephens2023}.}
    \label{tab:daniel_pars}
\end{table}

The scale height of the dust grains is the key parameter that will be tuned in our study. We allow each ring to have
different scale height profiles, parameterized as $H_{i}=H_{c,i}(R/C_i)^{p}$. In most cases, we let $p=1$,
describing a disk with a constant aspect ratio, $H/R$. In the gap, we allow the scale height to change from one
profile to another smoothly using a hyperbolic tangent function profile. Let $H_1(R)$ and $H_2(R)$ be the scale height profiles of the inner ring (disk) and the outer ring. Let $C$ and $W$ be the center and width of the gap, defined by the edges of the adjacent rings. We have:
\begin{equation}
\begin{split}
    H_{gap}(R) =& H_1(R) + 0.5[H_2(R)-H_1(R)]\times \\
    &\left[\tanh\left(\frac{R-C}{W/2}\right)+1\right].
\end{split}
\end{equation}

We set up our models using the Monte Carlo Radiation Transfer code, RADMC-3D\footnote{\url{https://www.ita.uni-heidelberg.de/~dullemond/software/radmc-3d/}}. 
The model is set in a spherical polar grid with $300\times 256\times 256$ cells in total. 
For each radiation transfer simulation, we use $1.6\times 10^9$ photons to obtain a polarization image with a high
signal-to-noise ratio. 
The dust opacities are calculated using the Mie theory code, \texttt{bhmie} \citep{BH83}. 
The results are smoothed by Gaussian beams before being compared with observations.
In the following subsections, the polarization features are studied on three different scales:
the inner disk ($\le 10$ au) in \ref{ssec:ID}, the first ring (ring 1$\sim 20$ au) in \ref{ssec:1R}, and
the outer disk (ring 2 and beyond, $>30$ au) in \ref{ssec:large_scale}.

\subsection{Inner disk: Near-far asymmetry}
\label{ssec:ID}
\begin{figure}
    \centering
    \includegraphics[width=\linewidth]{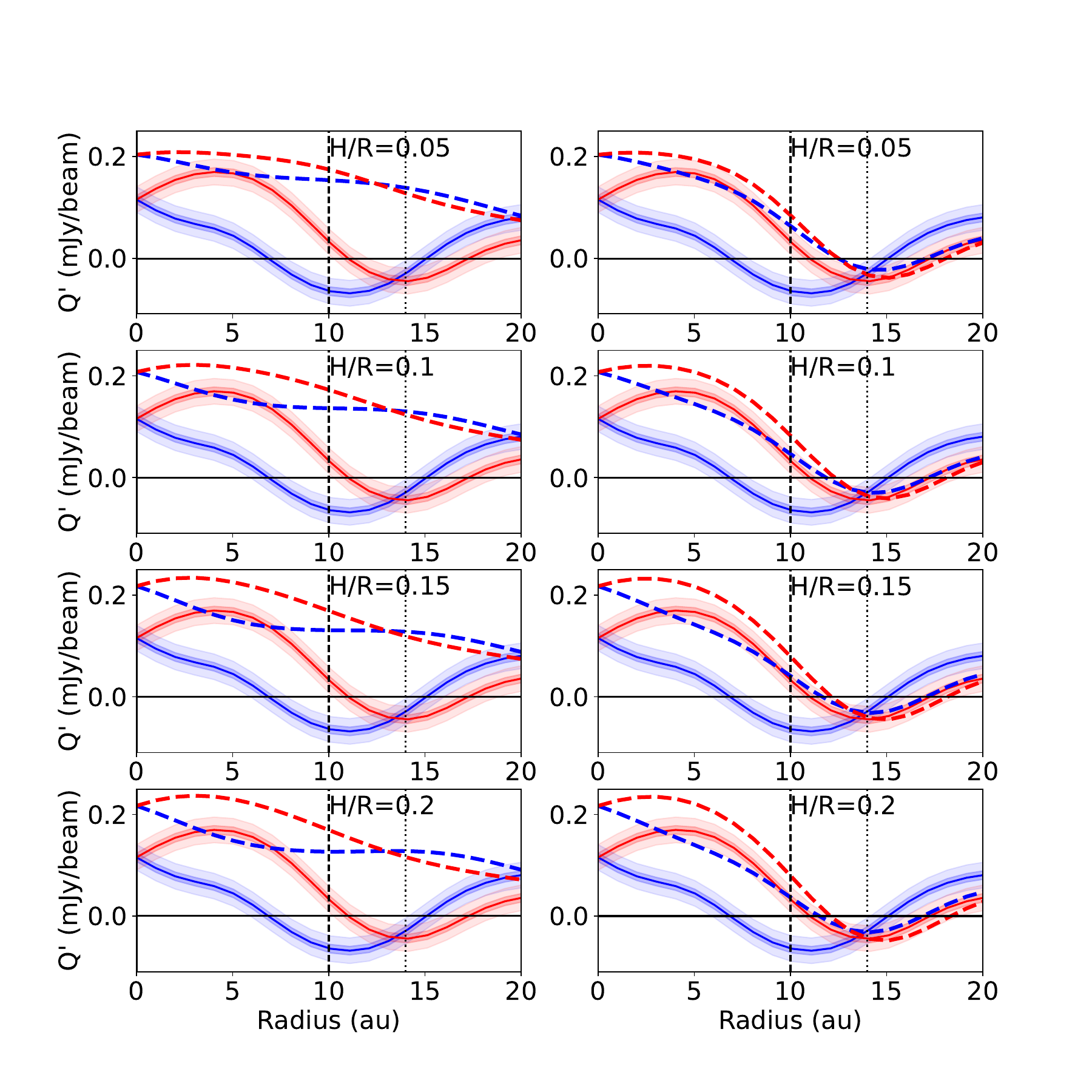}
    \caption{The $Q'$ profiles along the minor axes for the near and far sides of the inner disk, shown as red and blue curves, respectively. The solid lines represent the data, with darker and lighter shades represent $\sigma$ and $3\sigma$ errors, respectively. The dashed curves represent the models. 
    The vertical dashed and dotted lines represent the characteristic radius of the inner disk, $R_c$, and the location of the first gap, respectively.
    From the top row to the bottom row, we present models with $H/R=0.05,\, 0.1,\, 0.15,$ and $0.2$. The left column show the results from our models. The right column add a Gaussian profile with negative values at the location of the first gap and a width corresponding to the deprojected beam size along the minor axis. This extra Gaussian component is introduced to mimic the influence of the azimuthal polarization patterns in the gap. See text for more details. }
    \label{fig:ID}
\end{figure}

With the $\approx 5$ au beam size, we are able to resolve the inner disk with polarization observations.
Although the Stokes I image is very symmetric, the polarized intensity has a significant asymmetry between the near side (southwest) and the
far side (northeast) \footnote{According to the jet from the HL Tau system \citep{Mundt1988}, the northeast side is the far side and the southwest side is the near side.}.
The near side is much brighter than the far side (see the red curve and the blue curve in Figure~\ref{fig:ID}). 
Such an asymmetry is a natural outcome of a turbulent inner disk with a large aspect ratio \citep{Yang2017}. 
In this subsection, we study the dust scale heights in the inner disk and their effects on the
polarization features.

The left columns in Figure~\ref{fig:ID} shows the polarization profiles for models with $H/R=0.05, 0.1, 0.15, 0.2$. We can see that the near-far side asymmetry is below the detection limit for $H/R\le 0.05$, similar to what is observed at the large scale. For $H/R\ge 0.15$, the near-far side asymmetry is roughly saturated and does not increase any more. The maximum difference between the near side and the far side is comparable to the asymmetry in the observed data, although it is still slightly smaller than the observed one.  

Around $R_0=10$ au, the observed $Q'$ starts to become negative because the polarization orientation is in the azimuthal direction in the gap, perpendicular to the inclination-induced polarization from scattering.
This azimuthal pattern is believed to come from dichroic thermal emission from toroidally aligned effectively prolate grains (\citealt{Yang2019,Lin2022,Stephens2023}; see Section~\ref{ssec:residuals} for more discussions). 
In this paper, we focus on polarization from self-scattering of dust grains, but the contamination from
the first gap is non-negligible near the edge of the inner disk.
To mimic the transition of polarization patterns in the observed $Q'$ radial profile, we introduce a Gaussian profile with a peak flux density of $-0.16\rm\, mJy/beam$, centered at $14$ au, with a width of $5.1/(2\cos i)$ au, roughly the deprojected half-beam size along the minor axis, and $i=46.7^\circ$ is the inclination angle \citep{ALMA2015}. 
The results are shown in the right panels of Figure~\ref{fig:ID}. 

There are two caveats on the modeling of near-far side asymmetry at the inner disk.
On the one hand, the polarized intensity is higher than the observed one. This may be solved with different grain models, using either a smaller grain size or large and porous grain models. We will leave these explorations to future studies. 
On the other hand, $U'$ is not reproduced. The $U'$ from pure scattering is known to have different signs on the two sides with respect to the minor axis (see the upper right panel of Figure~\ref{fig:final_model}), known as the
bifurcation of polarization orientations \citep{Yang2017}. 
However, despite the fact that the observed $U'$ shows asymmetries, it does not have the predicted asymmetry with respect to the minor axis from our model.
It is likely that $U'$ is also substantially contaminated by the first gap. 
A quantitative and self-consistent modeling of these structures requires the proper treatment of scattering by aligned grains, ideally in Monte Carlo Radiation Transfer simulations. This approach has numerical difficulties and will be left for future exploration. 

\subsection{Ring 1: azimuthal contrast}
\label{ssec:1R}

\begin{figure*}[!htb]
    \centering
    \includegraphics[width=\linewidth]{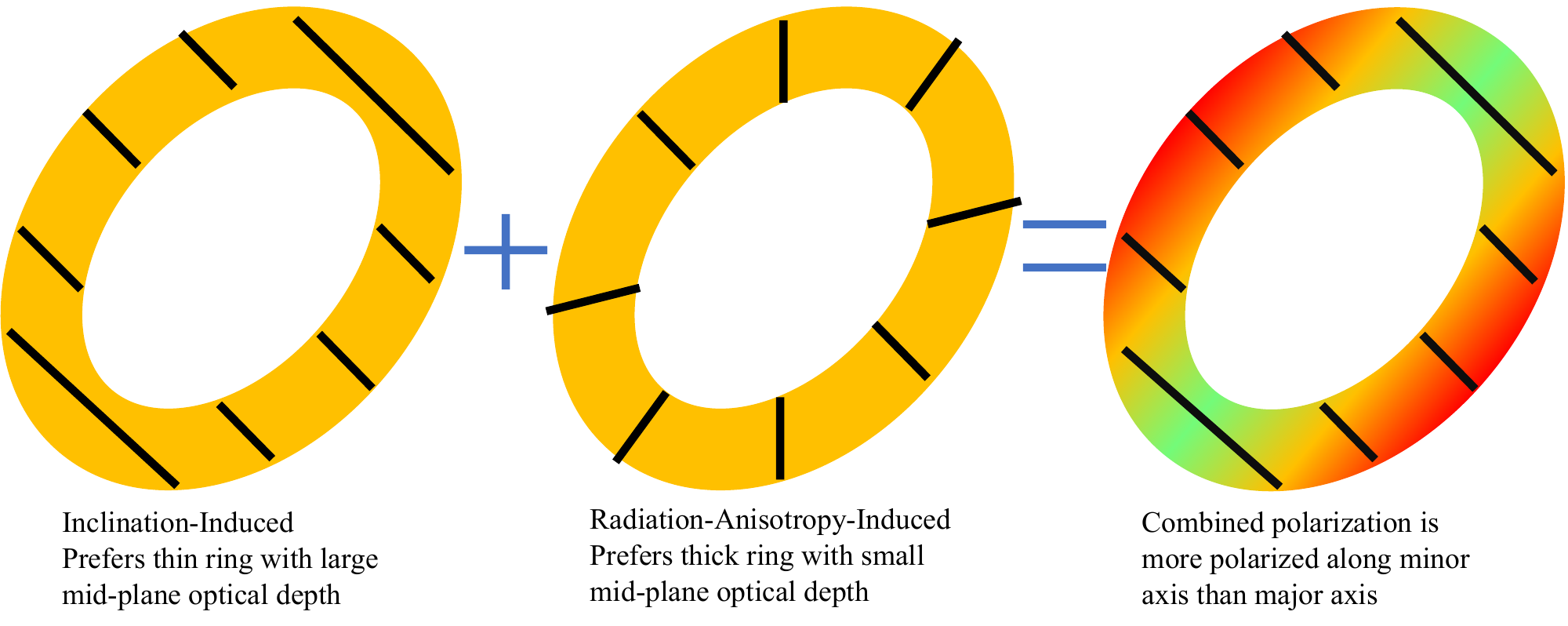}
    \caption{Visualization of how inclination-induced polarization and radiation-anisotropy-induced polarization contribute jointly to the observed polarization pattern.
    }
    \label{fig:illustration}
\end{figure*}

\begin{figure*}[!htb]
    \centering
    \includegraphics[width=0.8\linewidth]{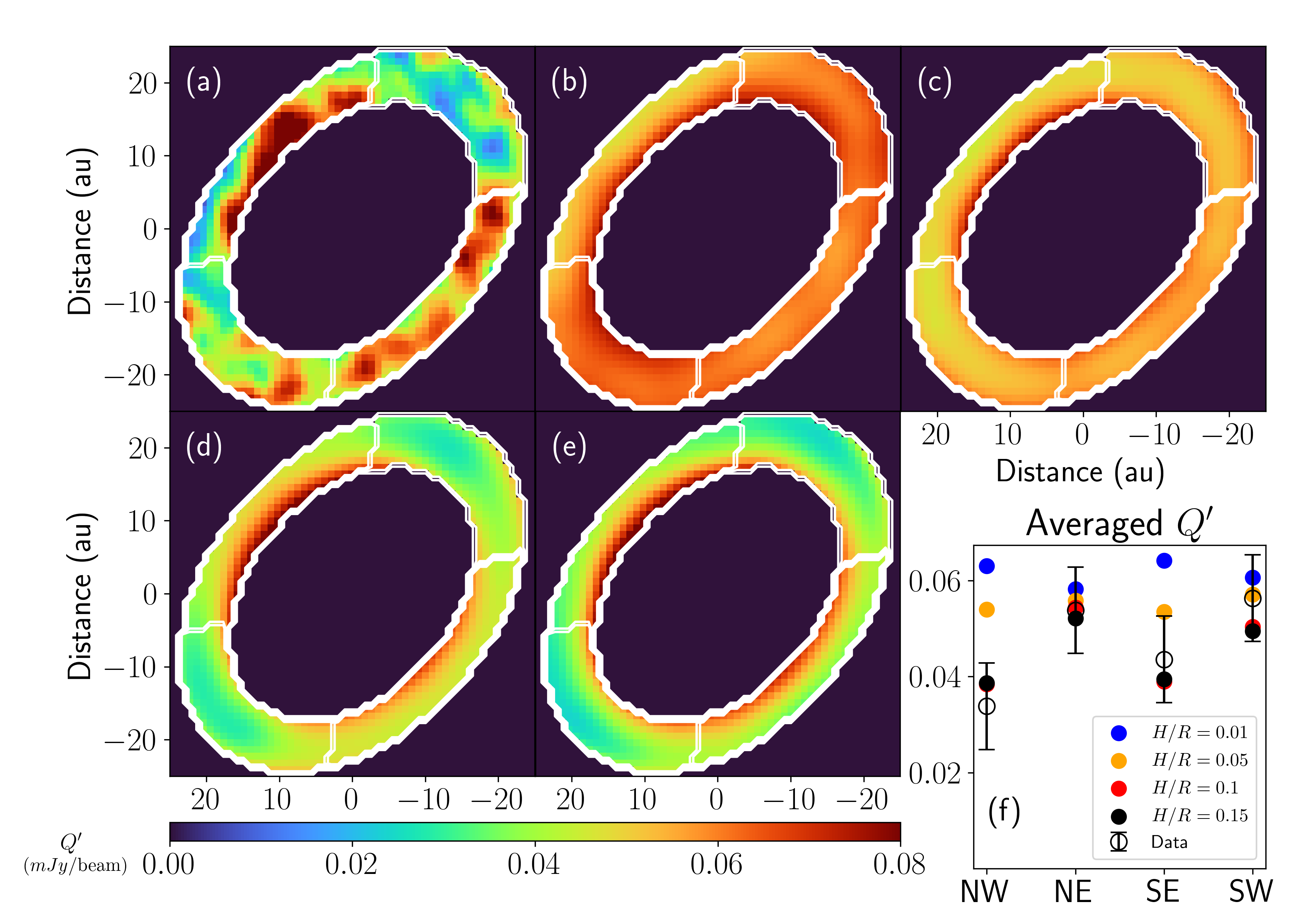}
    \caption{\textbf{Panels (a) - (e)}: The $Q'$ in the first ring. 
    \textbf{Panel (a)}: the data. 
    \textbf{Panels (b) - (e)}: Models 
    with $H/R=0.01,\, 0.05$, $0.1$, and $0.15$, respectively. 
    The white contours represent the four regions for averaging 
    the $Q'$. 
    \textbf{Panel (f)}: Averaged $Q'$ at the first ring. 
    }
    \label{fig:1R_image}
\end{figure*}

In the first ring ($\sim 20$ au), the polarization observations show azimuthal profiles that cannot be explained with the simplest
inclination-induced polarization models with settled dust grains. 
Specifically, continuum emission is more polarized in regions along the minor axis than in those along the major axis, 
shown in panel (a) of Figure~\ref{fig:1R_image}. 
For the inclination-induced polarization, the polarization fraction is uniform, to the leading order. 
In the presence of non-negligible radial radiation flux, the polarized intensity is usually slightly concentrated
along the major axis, i.e., more polarized along the major axis than the minor axis. 
This characteristic can be seen, for example, in the large-scale structures in Figure~\ref{fig:image}, or in the models of \cite{Yang2016a}. 
This is the opposite of what is observed in our data. 
As we show below, the observed pattern is a result of the
interplay between the inclination-induced polarization and the radiation-anisotropy-induced polarization (see Figure~\ref{fig:illustration}).

If the disk is infinitely thin, we expect the polarization fraction to be the same between the major and minor axes. 
This is the leading-order ``slab model'' in self-scattering, which produces a uniform polarization fraction and orientation \citep{Yang2017}.
This is also what we see in the model with $H/R=0.01$, Panel (b) in Figure~\ref{fig:1R_image}.
In this model, we also observe a slightly larger polarization flux along the major axis than along the minor axis.
This trend is similar to what is observed in the outer disk,
(Section~\ref{ssec:large_scale}), 
which is the result of some nonnegligible radial flux from the inner disk.

As we gradually increase the dust scale height, the optical depth along the disk midplane decreases. The scattering dust grains start to encounter the radiation fluxes across the ring-like structures. The self-scattering of dust grains in this ringlike structure is known to produce polarization perpendicular to the ring, roughly in the radial direction (see similar models in \citealt{Kataoka2015}). When the dust scale height is large, the polarization from such radiation anisotropy becomes more and more important. This pattern cancels the polarization at locations along the major axis, while adding to the polarization at locations along the minor axis. The result is a larger polarization along the minor axis than along the major axis, as shown in Panel (e) of Figure~\ref{fig:1R_image}. 

Note that in the data, the polarization at locations on the far side along the major axis is more polarized than that at locations
on the near side, which is the opposite of the well-studied near-far side asymmetry discussed by \cite{Yang2017}. 
This feature is also reproduced in the model with $H/R=0.1$. The reason is because in the first ring, the far side is slightly
brighter in total intensity Stokes $I$, so that even though the polarization fractions at these two locations are similar, the
polarized intensity is brighter in the far side along the major axis in the first ring. 

Panel (f) in Figure~\ref{fig:1R_image} shows the averaged $Q'$ in the four quarters of the ring for the data and for models with $H/R=0.01,0.05,0.1$. We can see that the model with $H/R=0.1$ best reproduces the azimuthal profile of $Q'$ in the first ring. We also experiment with models with $H/R=0.15,$ and 0.2. The results are similar to $H/R=0.1$. So, the aspect ratio at the first ring is $H/R\ge 0.1$ from our polarization modelings. This is in line with the constraint from the Band 9 continuum image \citep{Guerra-Alvarado2024}. 

\subsection{Outer disk: lack of near-far asymmetry}
\label{ssec:large_scale}

In contrast to the strong near-far side asymmetry on the inner disk, the polarization flux is relatively symmetric with respect
to the major axis on large scales, $R>30\rm\, au$. This means that the disk is likely flat with a small aspect ratio
at the outer disk. In this subsection, we try to quantify our constraints on the thickness of
the disk on large scales using the non-existence of near-far side asymmetry.

\begin{figure}
    \centering
    \includegraphics[width=\linewidth]{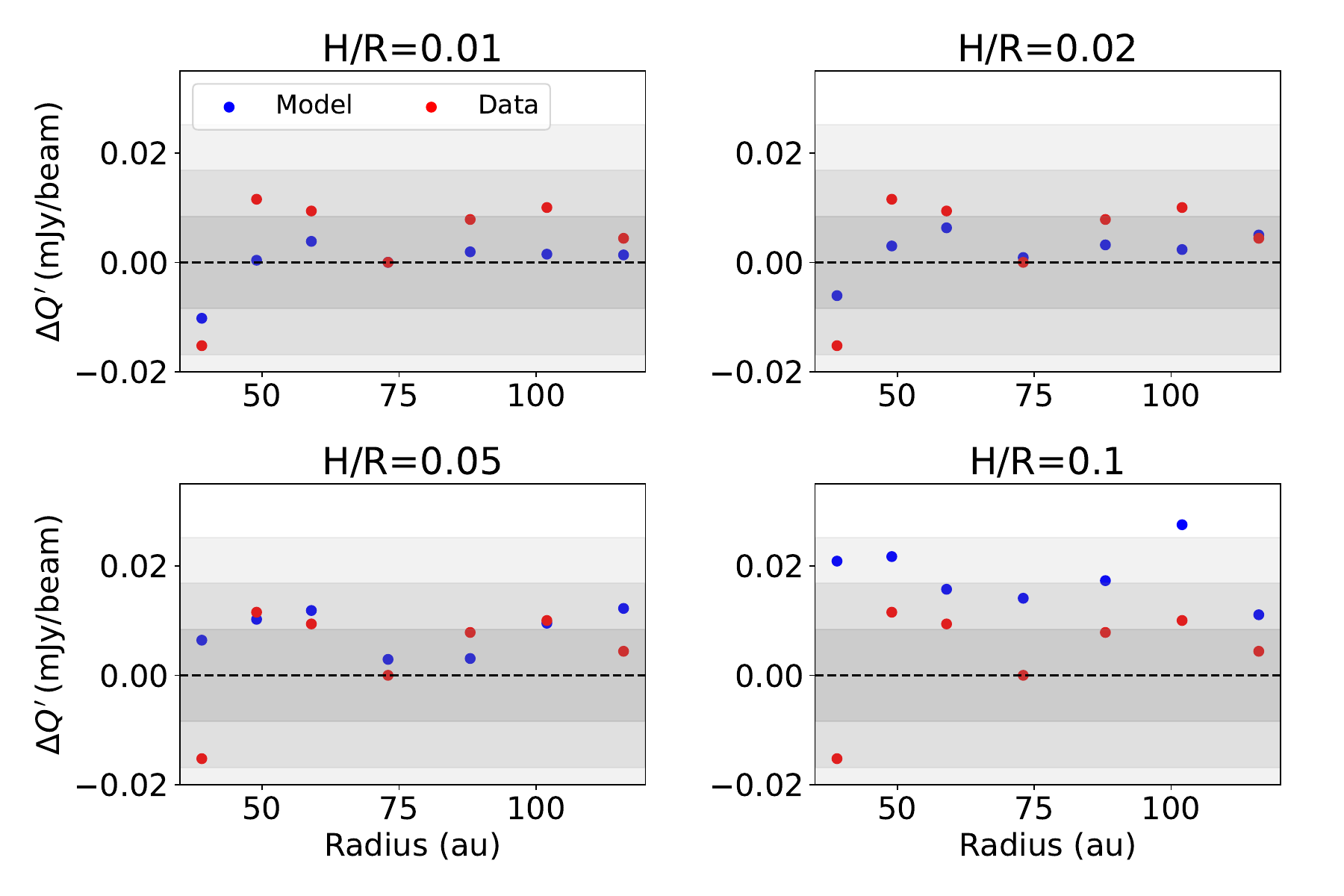}
    \caption{The difference of $Q'$ between the near side and the far side along the minor axis at each ring.
    Blue (Red) dots represent model (data). 
    The shades from the darkest to the lightest in the right panels represent $(1,2,3)\times \sigma_\mathrm{PI}$, respectively.}
    \label{fig:LS_Q_dQ}
\end{figure}

To compare the model with data more quantitatively, we average the $Q'$ over a 30-degree wedge and over the width of
each ring. We then subtract the far side $Q'$ from the near side value to obtain $\Delta Q'$ for each ring (shown in Figure~\ref{fig:LS_Q_dQ})
We can see that in the models with $H/R=0.1$, the near side is significantly brighter than the far side in
$Q'$. For models with $H/R<0.05$, the contrast is not significant. 
There is no definitive boundary as to exactly where the constraint is. We suggest that, roughly, the non-existence
of near-far side asymmetry at large scales yields a constraint of $H/R<0.05$.


Our constraint on the dust scale heights at the outer disk, $H/R<0.05$, is independent of the result of \cite{Pinte2016}, who constrained the scale height in the HL Tau disk to be approximately 0.7 au at a radius of 100 au,
which corresponds to $H/R=0.007$, less than the smallest aspect ratio in
our models. Polarization modeling is not sensitive to aspect ratios this small. In other models discussed in previous subsections, we adopt $H/R=0.007$ at 100 au. The exact choice does not affect the results at small scales, the inner disk and the first ring.
Our constraint is limited by the limited signal-to-noise ratio on large scales. 
It will be helpful if one can have deeper observations of the polarized intensity on large scales in the future, which may give us the ability to produce
constraints on a similar level using scattering-induced polarization.

\section{Discussion}
\label{sec:discussion}
\subsection{Scale height profile and turbulent stirring}
\label{ssec:HoR_prof}
In Section~\ref{sec:model}, we modeled the scattering-induced polarization and gives constraints on dust scale heights at the inner disk ($H/R\le 0.15$), the first ring ($H/R\le 0.1$), and the outer disk ($H/R\le 0.05$), separately. These constraints are plotted as gray-shaded regions in Figure~\ref{fig:hR}. 
We can see that the dust grains are much less settled in the inner part with larger scale height than those in the outer part. 
It cannot be reproduced with a constant turbulence $\alpha$ model. To model the dust scale height under the turbulence stirring model, we adopt a gas disk model for HL Tau \citep{Kwon2011}, which is a standard viscous accretion disk model \citep{Pringle1981}, as follows:

\begin{figure}[!htp]
    \centering
    \includegraphics[width=\linewidth]{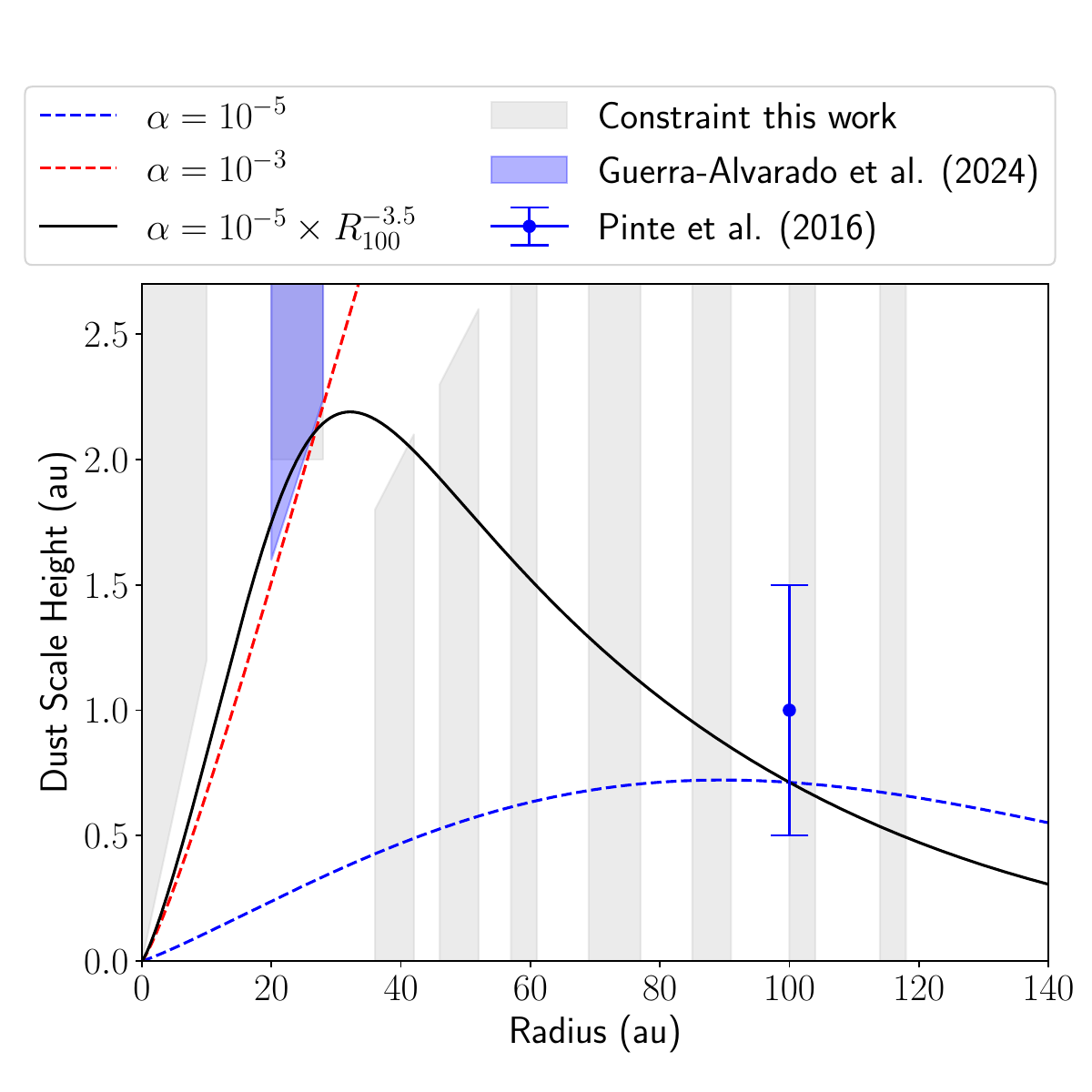}
    \caption{The constraints on scale heights and the scale height profiles for various models. 
    The grey shaded regions represent the suggested regions from our modeling on the polarization from dust
    self-scattering. 
    The dashed curves show two models with constant turbulence $\alpha$.
    The dot dashed curves show one model with varying turbulence $\alpha$. 
    The blue shaded region and the blue dot with a rough error bar represent the constraints from \cite{Guerra-Alvarado2024} and \cite{Pinte2016}, respectively.}
    \label{fig:hR}
\end{figure}



\begin{equation}
    \Sigma(R) = \Sigma_0 \left(\frac{R}{R_c}\right)^{-\gamma} e^{-(R/R_c)^{(2-\gamma)}},
\end{equation}
where $R_c=79$ au and $\gamma=0.2$. The total disk mass is $0.13\rm\, M_\odot$ \citep{Kwon2011}. 
The gas scale height profile, $H_g(R)$, can then be calculated using the central stellar mass 
$1.3\rm\, M_\odot$ \citep{ALMA2015}, the temperature profile in our model, and a mean 
molecular weight of $2.33$.

The degree of coupling of the dust grains with the gas is determined by the Stokes number $\mathrm{St}=\rho_s a /\Sigma_g$ and the turbulence $\alpha$, where we adopt $\rho_s=3\rm\, g/cm^3$ and $a=140\rm\, \mu m$. 
The dust scale height can be calculated as \citep{Youdin2007}:
\begin{equation}
    H_d = H_g \left(1+\frac{\mathrm{St}}{\alpha}\right)^{-1/2}\left(\frac{1+2\mathrm{St}}{1+\mathrm{St}}\right)^{-1/2},
    \label{eq:Hd}
\end{equation}

In Figure~\ref{fig:hR}, we plot two dashed curves representing models with constant turbulence $\alpha$ at levels of $10^{-5}$, and $10^{-3}$. We can see that to reproduce the dust scale height in the inner part of the disk, an $\alpha$ of $\sim 10^{-3}$ is required. 
However, this level of turbulence is too high for the dust grains settled on large scales. A model with constant turbulence cannot reproduce the high-resolution polarization observations in the HL Tau system. The inner part of the disk is more turbulent than the outer part.

To mimic the effect of different levels of turbulence, we draw one curve with varying $\alpha$ in Figure~\ref{fig:hR}, with $\alpha=10^{-5}\times R_{100}^{-3.5}$, where $R_{100}\equiv (R/100\,\mathrm{au})$. This model reproduces the constraints reasonably well, with $\alpha=10^{-5}$ at $100$ au and $\alpha=3\times 10^{-3}$ at the first ring.
The exact value of $\alpha$ and its profile are subject to the Stokes number, which
is determined by both the grain size and the gas column density. We would like to note that the
absolute values of $\alpha$ may scale as one changes the grain size and/or the disk mass: larger (smaller) grain size and/or smaller (larger) disk mass will result in a larger inferred $\alpha$. 
Different density profiles will also result in a contrast in $\alpha$ between the inner disk and the outer disk\footnote{Disk models for HL Tau in the literature all share similar radial power-law profiles. See, for example, a comparison in the upper left panel of Figure 3 in \cite{Pinte2016}.}, although it is likely a much higher order effect comparing with the large radial variation in $\alpha$ observed in our work.
With all these in mind, for a reasonable disk model, our main conclusion (the need for a change in levels of turbulence) is valid. 

The vertical dust scale height is crucial in the process of planet formation. 
The settling of dust grains into a thin vertical layer is the first step for enhancing the concentration of dust grains, which is prerequisite for the formation of planetesimals \citep{Youdin2007,Chiang2010}. 
From our constraint, the inner disk is thicker than the outer disk, 
which may implies small efficiency in the formation of planetesimals at 
$R< 30$ au. More studies are needed to understand the formation of planetesimals within such a thick layer of dust grains to put quantitative constraints on its impact on planet formation. 

The physical process driving such a large contrast in turbulence $\alpha$ is also puzzling. It is known that the inner disk is much more turbulent than the outer disk, due to the magneto-rotational instabilities inside the dead zone inner boundary. But that only happens very close to the central star when $R<1$ au, when the disk is hot enough to enable thermal ionization \citep{2021MNRAS.504..280J}.  
It is also possible that the large scale height is not due to a large turbulence in the disk midplane, but rather a result of disk winds, magnetically driven winds or photoevaporation winds \citep{Pascucci2023}. In any case, the dust scale height profile shown in Figure~\ref{fig:hR} is valuable to understand the dynamics of dust and gas in protoplanetary disks and important to the formation of planets.

\subsection{The polarization image in the final model and residual signals from grain alignment}
\label{ssec:residuals}

\begin{figure}
    \centering
    \includegraphics[width=0.5\textwidth]{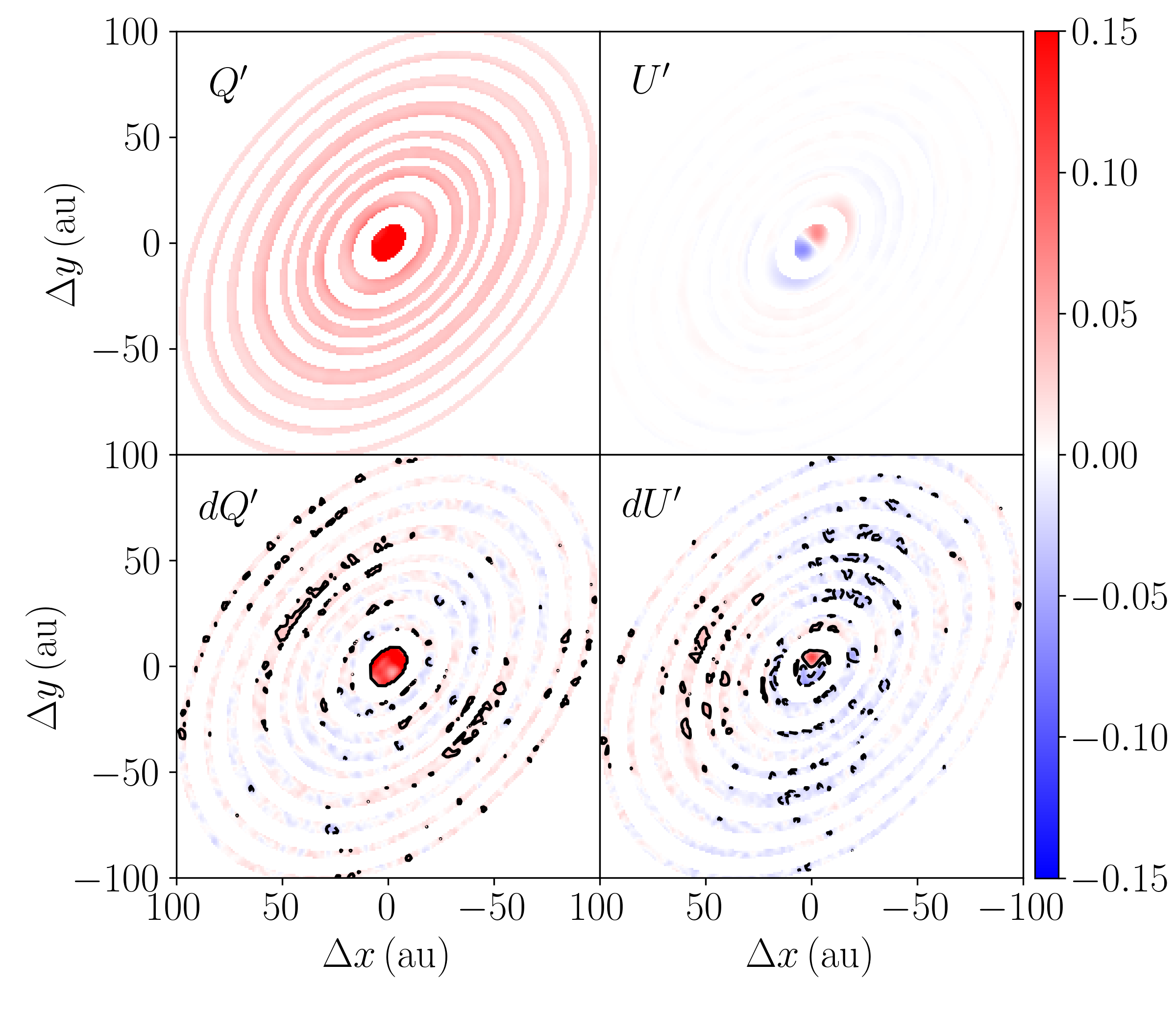}
    \caption{Comparison between our final model and the data. 
    \textbf{Upper panels}: The Stokes $Q'$ and $U'$ of the model. Note that the inner disk is saturated. 
    \textbf{Lower panels}: The difference between the Stokes parameters of the model and the data. The contours shows the $\pm3\sigma_\mathrm{Ip}$ contours. 
    All panels share the same colormap, which is shown on the right of the figure.
    The colormap is the same as the lower panels of Figure~\ref{fig:image}}
    \label{fig:final_model}.
\end{figure}

In Section~\ref{ssec:HoR_prof}, we put together scale height constraints from three different parts of the disk and define a scale height profile that roughly agrees with these constraints, shown as the black curve in Figure~\ref{fig:hR}. 
In Figure~\ref{fig:final_model}, we present the Stokes parameters in this model and the comparison with the observed data. 
The models are convolved with the observational synthesized Gaussian beam before comparing with the data.

We can see that it roughly agrees with the observations in the whole domain. 
In the inner disk, the polarization is higher than in the observation, which was discussed in Section~\ref{ssec:large_scale}. 

In an optically thick and geometrically thick disk, self-scattering polarization is known to produce a bifurcation of the polarization orientation at two sides of the disk with respect to the minor axis (see \citealt{Yang2017} and their Figure 4i). 
These structures are not obvious in the inner disk, probably because of the contamination from the first gap. 

We can also see that there are some systematic residuals in both $Q'$ and $U'$ in the outer disk. Specifically, the $Q'$ in the data is larger along the major axis than the minor axis, and the $U'$ has some quadruple modes. Qualitatively, these residuals may be signatures of dichroic emission from aligned dust grains. They can be either toroidally aligned prolate grains or radially aligned oblate grains. Currently, toroidally aligned prolate grains are the preferred model \citep{Yang2019}. 
\cite{Lin2024b} fitted the grain alignment components in multiple bands for the HL Tau system. 
On the theory side, recently, \cite{Thang2024} proposed that such signals may also come from grains aligned with toroidal magnetic fields but with a wrong internal alignment.
\cite{Lin2024b} studied the dynamics of a dust grain with the center of mass shifted from the geometric center under a stream of gas fluid. They found that the dust grains will be aligned with the shifting vector being aligned along the differential motion between the gas and the dust grains, supporting the picture of toroidally aligned prolate grains.
We will leave quantitative studies on these residuals for future study. 

\section{Summary}
\label{sec:summary}

In this work, we model the high-resolution polarization observations towards the HL Tau system, focusing 
on the polarization from dust self scattering that dominates the inner disk and rings. The results are summarized as follows.

\begin{enumerate}
    \item The high-resolution polarization observations reveal a substantially asymmetric inner disk in the polarized flux; specifically, the near side is more polarized than the far side of the inner disk. This near-far side asymmetry is in agreement with the prediction of \cite{Yang2017}. 
    \item We modeled the observed near-far side asymmetry in the inner disk with varying aspect ratios $H/R$. We find that this asymmetry increases with increasing $H/R$. It roughly saturates after $H/R\ge 0.15$, which best reproduces the observed asymmetry. We conclude that the inner dust disk has a large aspect ratio of $H/R\ge 0.15$.
    \item In the first ring, we observe an azimuthal contrast with the polarized flux being larger along the minor axis than the major axis, which is indicative of a moderately thick dust ring. This contrast is attributed to the interplay between the inclination-induced polarization and the polarization from radiation anisotropy. The radiation anisotropy is influenced by the optical depth along the disk midplane, which is in turn controlled by the dust scale height. This phenomenon is analogous to the findings of \cite{Yang2024}. 
    Through a series of modeling with varying dust aspect ratio focusing on the first ring polarization,
    we find that the observed polarization contrast is best reproduced with an aspect ratio of $H/R\approx 0.1$ or larger.
    \item Putting all of our constraints on the dust scale heights at different scales together, we conclude that the inner disk is more turbulent than the outer disk. 
    Models with constant turbulence $\alpha$ cannot reproduce the scale height profile (see Figure~\ref{fig:hR}). 
    A model with varying $\alpha$ that goes from $10^{-5}$ at $100$ au to $10^{-2.5}$ at $20$ au is most consistent with all the constraints. 
    \item There are some small yet systematic residuals that may arise from toroidally oriented prolate grains \citep{Yang2019,Stephens2023}. 
    Further detailed analysis is needed to confirm and analyze the strength of such signals. 
\end{enumerate}

This paper makes use of the following ALMA data: ADS/JAO.ALMA$\#$2019.1.01051.S and 
ADS/JAO.ALMA$\#$2016.1.00115.S. ALMA is a partnership of ESO (representing its member states), NSF (USA) and NINS (Japan), together with NRC (Canada), NSTC and ASIAA (Taiwan), and KASI (Republic of Korea), in cooperation with the Republic of Chile. The Joint ALMA Observatory is operated by ESO, AUI/NRAO and NAOJ.

H.Y. is partially supported by National Natural Science Foundation of China (NSFC) [12473067] and Tang Foundation.
Z.Y.L. is supported in part by NASA 80NSSC20K0533 and NSF AST-2307199.
L.W.L. acknowledges support from NSF AST-1910364 and NSF AST-2307844.

\bibliographystyle{aasjournal}



\end{CJK*}
\end{document}